\documentclass[a4paper,12pt,doublespace]{article}
\usepackage[utf8x]{inputenc}

\usepackage{dcolumn}

\newcolumntype{.}{D{.}{.}{4}}

\usepackage{setspace}
\usepackage{chngcntr}

\usepackage[left=2cm,top=2cm,right=2cm,bottom=2cm]{geometry}

\usepackage{rotating}
\usepackage{graphicx}
\usepackage{color}

\usepackage{xcolor}
\usepackage{amsmath}
\usepackage{amsfonts}
\usepackage{amssymb}
\usepackage{soul}
\usepackage{bbold}
\usepackage{subfig}

\usepackage[round]{natbib}

\begin{document}
\raggedright

\centering
\Large{Improved Dynamic Predictions from Joint Models of Longitudinal and Survival Data with Time-Varying Effects using P-splines}

\vspace{10 mm}

\centering
\large{Eleni-Rosalina Andrinopoulou$^1$, Paul H.C. Eilers$^1$, Johanna J.M. Takkenberg$^2$ and Dimitris Rizopoulos$^1$}

\vspace{10 mm}

\normalsize{
1. Department of Biostatistics, Erasmus MC, Rotterdam, The Netherlands\\
2. Departments of Cardiothoracic Surgery, Erasmus MC, Rotterdam, The Netherlands}

\vspace{10 mm}

\normalsize{Corresponding author: Eleni-Rosalina Andrinopoulou, Department of Biostatistics, Erasmus MC, PO Box 2040, 3000 CA Rotterdam, The Netherlands\\
email: e.andrinopoulou@erasmusmc.nl, Tel: +31/10/7043731, Fax: +31/10/7043014}

\singlespacing
\abstract{
In the field of cardio-thoracic surgery, valve function is monitored over time after surgery. The motivation for our research comes from a study which includes patients who received a human tissue valve in the aortic position. These patients are followed prospectively over time by standardized echocardiographic assessment of valve function. Loss of follow-up could be caused by valve intervention or the death of the patient. One of the main characteristics of the human valve is that its durability is limited. Therefore, it is of interest to obtain a prognostic model in order for the physicians to scan trends in valve function over time and plan their next intervention, accounting for the characteristics of the data.

Several authors have focused on deriving predictions under the standard joint modeling of longitudinal and survival data framework that assumes a constant effect for the coefficient that links the longitudinal and survival outcomes. However, in our case this may be a restrictive assumption. Since the valve degenerates, the association between the biomarker with survival may change over time.

To improve dynamic predictions we propose a Bayesian joint model that allows a time-varying coefficient to link the longitudinal and the survival processes, using P-splines. We evaluate the performance of the model in terms of discrimination and calibration, while accounting for censoring.
}

\newpage

\doublespacing
\section{Introduction}\label{Intro}
In the field of cardio-thoracic surgery, valve function is monitored periodically over time after heart valve surgery. Aortic gradient (AG) (mmHg) is one of the continuous echocardiographic markers that measures valve (dys)function, where high values indicate a worsening of the patient's condition. Specifically, it measures aortic stenosis which occurs when the opening of the aortic valve located between the left ventricle of the heart and the aorta is narrowed. During the follow-up period after surgery, patients may require an intervention or may die. The motivation of this research comes from a study, conducted in the Erasmus University Medical Center, which includes all patients who received a human tissue valve allograft in the aortic position in the Department of Cardio-Thoracic Surgery \citep{bekkers2011re}. In total 296 patients who survived aortic valve or root replacement with an allograft valve were followed over time. Specifically, echocardiographic examinations were scheduled at six months, one year postoperatively and biennially thereafter. During follow-up, 161 (54\%) patients either died or required a reoperation on the same valve. A total of 1669 echocardiographic measurements of AG were performed. The median number of visits is six and the median years of follow up is 9.3. One of the characteristics of human valves is that their durability is limited. Hence, it is important for the physicians to have a prognostic tool in order to carefully monitor trends in valve function over time and plan a future re-intervention.

\begin{figure}[h!]
\centerline{%
\includegraphics[width=5in]{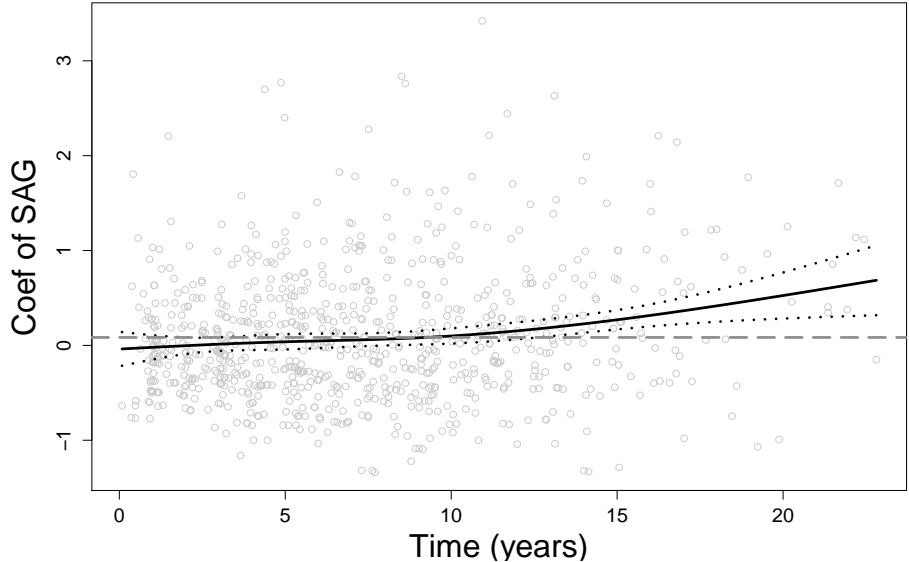}}
\caption{Time-varying coefficient of the SAG, using the Schoenfeld residuals. The solid line represents the mean estimate while the dotted lines the corresponding confidence interval. The dashed grey line represents the coefficient of the SAG.}
\label{SchRes}
\end{figure}

Joint modeling of longitudinal and survival data is a popular framework to analyze data including repeated measurements and time-to-event outcomes appropriately \citep{tsiatis2004joint,rizopoulos2012joint,andrinopoulou2012introduction, rizopoulos2015personalized}. The idea behind these models is that the longitudinal and the survival processes share common random effects, inducing correlation between the two processes. Specifically, we construct a mixed-effects model to describe the evolution over time for the longitudinal outcome, and use these estimated evolutions as a time-dependent covariate in a survival model. Several authors have focused on deriving predictions under joint modeling of longitudinal and survival data framework (\citealp*{taylor2005individualized}; \citealp{garre2008joint}; \citealp*{yu2008individual}; \citealp{proust2009development}; \citealp{rizopoulos2011dynamic}; \citealp{andrinopoulou2015dynamic}). To improve the fit and the predictive accuracy of joint models, previous work allowed the inclusion of multiple longitudinal outcomes and investigated the selection of the optional functional form \citep{rizopoulos2011dynamic, rizopoulos2014combining, andrinopoulou2015combined, andrinopoulou2016bayesian}. A common feature of all aforementioned models and previous work published on the motivating data is that the parameters that measure the strength of the association between the longitudinal and survival outcome were assumed to be constant in time. However, the heart valve degenerates, therefore it is natural and biologically more desirable to assume that the effect of AG is changing over time. To investigate that, we performed a preliminary analysis assuming a time-dependent Cox model for the composite event death/reoperation including as predictors the square root of AG (SAG), a transformation that was explored in previous work \citep{andrinopoulou2014joint}, and gender. Figure~\ref{SchRes} shows the smoothed scatterplot of the Schoenfeld residuals from this Cox model versus time. These residuals are typically used to investigate the proportional hazards assumption, and for the SAG they show an increasing trend indicating violation of a constant-effect assumption.

Even though the consideration of time-varying coefficients has been extensively studied in the general context of survival analysis using polynomials and B-splines \citep{nan2005varying, perperoglou2014cox}, relatively little work has been done on joint models with time-varying coefficients \citep{song2008semiparametric}. To our knowledge, no work has been done to evaluate whether such time-varying coefficients may improve the accuracy of individualized predictions within the framework of joint models. The idea behind the time-varying coefficient models is to include interactions of the covariates with an appropriate pre-defined time function. To enhance the predictive performance of the joint model, we assume a time-varying effect of SAG using P-splines. Specifically, it is approximated by a polynomial spline written in terms of a linear combination of B-spline basis functions (\citealp{eliers1996flexible}; \citealp{lang2004bayesian}; \citealp*{eilers2015twenty}). To overcome the problem of the large number of parameters and to stabilize the predictions, a penalty is applied to the coefficients. To facilitate flexible modeling of the survival outcome, we use P-splines also for the logarithm of the baseline hazard. To evaluate the derived predictions we present extensions of classic measures of predictive ability, such as discrimination and calibration, in the time-dependent setting while accounting for censoring.

The rest of the paper is organized as follows. Section 2 describes the formulation of the joint model. Section 3 presents the Bayesian estimation. Section 4 presents measures to assess the predictive performance of the model. Section 5 shows the results for the cardio data analysis, while Section 6 contains simulation studies. Finally, in Section 7 we close with a discussion.

\section{Joint Model Definition}\label{JM}
We let $\mathrm{\textit{T}}_{i}^*$ denote the true failure time for the $i$-th individual ($i = 1,\dots,n$), and $C_i$ the censoring time. Moreover, $T_i=\min(\mathrm{\textit{T}}_{i}^*,C_i)$ denotes the observed failure time and $\delta_i = \{0, 1\}$ is the event indicator where zero indicates censoring. We let $\boldsymbol{y_{i}}$ denote the longitudinal response obtained at different time points $t_{ij}>0$, $(j = 1, \ldots, n_i)$. To describe the subject-specific evolution over time of the continuous longitudinal outcome SAG, we use a mixed-effects model. In particular, we postulate
\[
y_{i}(t) = \eta_i(t) + \epsilon_i = \boldsymbol{x}^\top_{i}(t)\boldsymbol{\beta} + \boldsymbol{z}^\top_{i}(t)\boldsymbol{b}_{i} + \epsilon_i(t),
\]
where $\boldsymbol{x}_{i}(t)$ denotes the design vector for the fixed effects regression coefficients $\boldsymbol{\beta}$ and $\boldsymbol{z}_{i}(t)$ the design vector for the random effects $\boldsymbol{b}_{i}$. Moreover, $\epsilon_i(t) \sim N(0, \sigma)$. For the corresponding random effects, we assume a multivariate normal distribution, namely
\[
\boldsymbol{b}_i \sim N_b(\boldsymbol{0},\boldsymbol{\Sigma}_b^2).
\]

We postulate a varying-coefficient joint model (VCJM) for the relationship between the survival and the longitudinal outcome. Specifically, we have
\[
h_{i}(t,\boldsymbol{\theta}_s)=h_{0}(t)\exp [\boldsymbol{\gamma}^{\top} \boldsymbol{w}_{i}+ f\{\lambda(t), \mathcal H_{i}(t)\} ],
\]
where $\boldsymbol{\theta}_s$ is the parameter vector for the survival outcomes, $\boldsymbol{w}_{i}$ is a vector of baseline covariates with a corresponding vector of regression coefficients $\boldsymbol{\gamma}$, $\boldsymbol{\gamma}_{h_0}$ is the vector of the baseline hazard coefficient and $\mathcal H_{i}(t) = \{\eta_{i}(\zeta),\ 0\leq \zeta <t\}$ denotes the history of the true unobserved longitudinal process up to time point $t$. The function $f\{\lambda(t), \mathcal H_{i}(t)\}$ specifies which features of the longitudinal submodel are included in the relative risk model \citep{rizopoulos2011bayesian,rizopoulos2012joint}. Previous work suggested that not only the value of SAG but also other characteristics of the biomarker may have an influence on the event of interest \citep{andrinopoulou2015combined}. Several specifications of $f(.)$ have been proposed in the literature \citep{rizopoulos2011bayesian}. Some examples are the following:
\[
\begin{array}{ll}%
f\{\lambda(t),  \mathcal H_{i}(t)\} = \lambda(t)\eta_i(t),\\
f\{\lambda(t),  \mathcal H_{i}(t)\} = \lambda_1(t)\eta_i(t) + \lambda_2(t)\frac{\textrm{d}\eta_i(t)}{\textrm{d}t},\\
f\{\lambda(t),  \mathcal H_{i}(t)\} = \lambda(t)\int_0^t \eta_i(s)ds.
\end{array}
\]
Specifically, $f(.)$ postulates that the hazard of the event is associated with the underlying value of the longitudinal outcome at a specific time point $t$, the value and the slope of the longitudinal outcome at $t$, or the accumulated longitudinal process up to time $t$.

In the standard constant-coefficient joint model (CCJM), $\lambda(t)$ is assumed to be constant over time. For the VCJM, we model $\lambda(t)$ flexibly assuming a smooth function. Several approaches have been proposed for modelling and estimating smooth functions such as B-splines. A problem that arises in such methods is the selection of the number and position of the knots which has been a subject of much research. In this manuscript, we adopt the P-splines approach for $\lambda(t)$. The basic idea of P-splines is to use a (relatively) high number of equally spaced knots. To obtain sufficient smoothness of the fitted curves and to avoid overfitting, a roughness penalty based on differences of adjacent B-spline coefficients is applied \citep{eliers1996flexible}. In particular, we take
\[
\lambda(t) = \sum_{\ell=1}^{L} \alpha_{\ell} B_\ell(t),
\]
where $\alpha_{\ell}$ is a set of parameters that capture the strength of association between the longitudinal and survival outcomes and $B_\ell(t)$ denotes the $\ell$-th basis function of a B-spline. To further facilitate improving the derived predictions from the joint model, we model the baseline hazard with the same P-splines approach. Specifically,
\[
\log\{h_{0}(t)\}=\sum_{u=1}^{U}\gamma_{h_0,u}B_u(t),
\]
where $\gamma_{h_0,u}$ are the coefficients of the baseline hazard and $B_u(t)$ denotes the $u$-th basis function of a B-spline.

\section{Bayesian Estimation}\label{Est}
We employ a Bayesian approach where inference is based on the posterior of the model. In particular, we use Markov chain Monte Carlo (MCMC) methods to estimate the parameters of the VCJM. The likelihood of the model is derived under the assumption that the longitudinal and survival processes are independent given the random effects  \citep{rizopoulos2012joint}. Moreover, the longitudinal responses of each subject are assumed independent given the random effects. The posterior distribution is written as
\[
p(\boldsymbol{\theta} \mid \boldsymbol{y}_{i},T_i,\delta_i)\propto  \prod_{j=1}^{n_{i}}p(y_{ij}\mid \boldsymbol{b}_{i},\boldsymbol{\theta}_y)  p\{T_i,\delta_i\mid \mathcal \eta_i(T_i),\boldsymbol{\theta}_s\}p(\boldsymbol{b}_{i}\mid \boldsymbol{\theta}_{y}) p(\boldsymbol{\theta}_{y}) p(\boldsymbol{\theta}_s),
\]
where $\boldsymbol{\theta} = (\boldsymbol{\theta}_s^{\top}, \boldsymbol{\theta}_{y}^{\top})^{\top}$ is the parameter vector for the survival and the longitudinal outcomes respectively. The likelihood contribution of the longitudinal and survival model together with the formulation of the deviance information criterion (DIC) are given in the Appendix.

\subsection{Bayesian P-splines}
The Bayesian P-splines approach was first introduced by \cite{lang2004bayesian}. In our case, the smoothness of functions $\lambda(t)$ and $h_0(t)$ is controlled by the following priors for the coefficient that links the longitudinal and the survival outcomes $\boldsymbol{\alpha}$ and the coefficient of the baseline hazard $\boldsymbol{\gamma}_{h_0}$:
\[
\begin{array}{ll}%
\boldsymbol{\alpha}\mid \tau_{\alpha} &\sim N_L(\boldsymbol{0}, \tau_{\alpha} \boldsymbol{M}_{\alpha} )\ \ \textnormal{and} \ \ \tau_{\alpha} \sim Gamma(c_1, c_2),\\
\boldsymbol{\gamma}_{h_0}  \mid \tau_{\gamma_{h_0}} &\sim N_U(\boldsymbol{0}, \tau_{\gamma_{h_0}} \boldsymbol{M}_{\gamma_{h_0}} ) \ \ \textnormal{and} \ \ \tau_{\gamma_{h_0}} \sim Gamma(f_1, f_2),
\end{array}
\]
where $\boldsymbol{M}_{\alpha}$, $\boldsymbol{M}_{\gamma_{h_0}}$ are the penalty matrices. In particular, $\boldsymbol{M}_{\alpha} = \boldsymbol{M}_{\gamma_{h_0}} = \boldsymbol{\mathcal{D}}_r^{\top}\boldsymbol{\mathcal{D}}_r + 10^{-6} \boldsymbol{I}$, where $\boldsymbol{\mathcal{D}}_r$ is a $r$-th order difference matrix. The scaled identity matrix $\boldsymbol{I}$ ensures a positive define variance-covariance matrix. As described in the literature \citep{reinsch1967smoothing,eliers1996flexible}, a common choice for the penalty matrices is to assume a second order penalty. The amount of smoothness is controlled by the variance parameters $\tau_{\gamma_{h_0}}$ and $\tau_{\alpha}$, where hyperpriors are assigned. A usual recommendation is to set $c_1$ and $f_1$ equal to 1 and $c_2$ and $f_2$ equal to a small number. Alternative specifications of these hyperpriors can be found in \cite{jullion2007robust}.

\section{Measuring Predictive Performance}\label{predPerf}
As motivated in Section~\ref{Intro}, it is important for physicians to have a prognostic tool for planning next interventions. To assess the predictive performance of the VCJM and to compare it to the CCJM, we focus on discrimination and calibration. Specifically, discrimination is how well can the model discriminate between patients who will experience the event from patients who will not \citep{pencina2008evaluating}, whereas calibration is how well the model predicts the observed event rates \citep{schemper2000predictive}.

\subsection{Discrimination}
A key feature of our model is to distinguish between patients who are going to die or require a reoperation within a specific time frame from patients who will not. In particular, for a future patient $l$ with SAG measurements $\boldsymbol{\tilde{y}_{l}}$ up to time point $t$, we are interested in investigating whether he will die or require a reoperation in the medically-relevant time frame $(t, t + \Delta t]$ within which the physician could intervene to improve survival. The survival/intervention-free probability of patient $l$ within this interval is,
\[
\pi_{l}(t, \Delta t)=\mbox{Pr}(T^*_{l}\geq  t+\Delta t \mid T^*_{l}> t, \boldsymbol{\tilde{y}_{l}(t)},\boldsymbol{D}_n),
\]
where $\boldsymbol{D}_n = \{T_{i}, \delta_i, \boldsymbol{y}_{i}, i = 1, \ldots, n\}$ denotes the sample on which the joint model was fitted. For the calculation of sensitivity and specificity, we have $\pi_l(t, \Delta t) \leq c$ if subject $l$ died or required a reoperation and $\pi_l(t, \Delta t) \geq c$ if he did not experience death/reoperation, for a specific $c\in[0, 1]$. In particular, we can define sensitivity and specificity as
\[
\mbox{Pr}\{ \pi_l(t, \Delta t) \leq c \mid T_l^* \in (t,t + \Delta t] \} \ \ \textnormal{and} \ \ \mbox{Pr}\{ \pi_l(t, \Delta t) > c \mid T_l^* > t + \Delta t \},
\]
respectively. Using the area under the receiver operating characteristic curve (AUC) we can assess the discriminative capability of the model. In particular, given a randomly chosen pair of patients $(l_1, l_2)$, we have
\[
\mbox{AUC}(t,\Delta t)=
\mbox{Pr}[\pi_{l_1}(t, \Delta t) < \pi_{l_2}(t, \Delta t) \mid \{ T_{l_1}^* \in (t,t + \Delta t]\} \cap   \{T_{l_2}^* > t + \Delta t\}  ].
\]
If patient $l_1$ experiences death/reoperation within the relevant time frame whereas patient $l_2$ does not, then we would expect the VCJM to assign higher survival/intervention-free probability during the period $(t, t + \Delta t]$ for the patient that did not experience death/reoperation. In the cardio data set, the values of time-to-death/reoperation are not fully observed for all patients. To account for this, the estimation of AUC$(t,\Delta t)$ is based on the following decomposition
\[
\widehat{\mbox{AUC}}(t,\Delta t) = \sum_{w=1}^4\widehat{\mbox{AUC}}_w(t,\Delta t),
\]
which include the following pairs of patients
\[
\begin{array}{ll}%
\Omega_{l_1l_2}^{(1)}(t) &=  [  \{  T_{l_1} \in (t, t + \Delta t]  \}  \cap \{\delta_{l_1} = 1 \}  ] \cap  \{ T_{l_2} > t + \Delta t \}, \\
\Omega_{l_1l_2}^{(2)}(t) &= [  \{  T_{l_1} \in (t, t + \Delta t]  \}  \cap \{\delta_{l_1} = 0 \}  ] \cap \{ T_{l_2} > t + \Delta t \},\\
\Omega_{l_1l_2}^{(3)}(t) &= [  \{  T_{l_1} \in (t, t + \Delta t]  \}  \cap \{\delta_{l_1} = 1 \}  ] \cap [ \{ T_{l_1} < T_{l_2} \leq t + \Delta t \}  \cap \{\delta_{l_2} = 0 \} ], \\
\Omega_{l_1l_2}^{(4)}(t) &= [  \{  T_{l_1} \in (t, t + \Delta t]  \}  \cap \{\delta_{l_1} = 0 \}  ] \cap [ \{  T_{l_1} < T_{l_2} \leq t + \Delta t  \}  \cap \{\delta_{l_2} = 0 \} ]. \\
\end{array}
\]
$\widehat{\mbox{AUC}}_1(t,\Delta t)$ includes the pairs of patients who are comparable $\Omega_{l_1l_2}^{(1)}(t)$ and $\sum_{w=2}^4\widehat{\mbox{AUC}}_w(t, \Delta t)$ the pairs of patients who due to censoring cannot be compared \{$\Omega_{l_1l_2}^{(w)}(t)$, $w = 2, 3, 4$\}. Then, with $I(.)$ the indicator function
\[
\widehat{\mbox{AUC}}_w(t,\Delta t) = \frac{\sum_{l_1=1}^{n} \sum_{l_2=1;l_2\neq l_1}^{n} I\{ \hat{\pi}_{l_1}(t, \Delta t) < \hat{\pi}_{l_2}(t, \Delta t) \} \times I\{\Omega_{l_1l_2}^{(w)}(t) \} \times \hat{K_w} }   {  \sum_{l_1=1}^{n} \sum_{l_2=1;l_2\neq l_1}^{n}  I\{\Omega_{l_1l_2}^{(w)}(t) \}\times \hat{K_w} },
\]
which is the proportion of concordant subjects out of the set of comparable subjects at time $t$. Specifically, if in a randomly selected pair of patients, the one with the higher event probability experiences the event and the one with the lower probability does not experience the event, then this pair is said to be a concordant pair. For $w = 1$, we have $\hat{K_2} = 1$ because the pairs of patients are comparable. For $w = 2,\ 3,\ 4$, the $\widehat{\mbox{AUC}}_w(t,\Delta t)$ are weighted with the probability that the concordant subjects are comparable. In particular, $\hat{K_2} = 1 - \hat{\pi}_{l_1}(t, \Delta t)$, $\hat{K_3} = \hat{\pi}_{l_2}(t, \Delta t)$ and $\hat{K_4} = \{1 - \hat{\pi}_{l_1}(t, \Delta t)\}\times  \hat{\pi}_{l_2}(t, \Delta t)$

\subsection{Calibration}
To assess the accuracy of the model, we use the prediction error (PE). Using all available information for a particular patient $l$, we are interested in comparing the predicted probability of survival/intervention-free of this patient to the observed truth:
\[
\mbox{PE}(t, \Delta t) = E[ \{ N_l(t + \Delta t) - \pi_l(t, \Delta t)  \}^2  ]
\]
where $N_l(t) = I(T_l^* >t)$ is the event status at time $t$. To account for censoring, the following estimate has been proposed by \cite{henderson2002identification}:
\[
\begin{array}{ll}%
\lefteqn{\widehat{\mbox{PE}}(t, \Delta t) =  } \\
\{ \mathcal{R}(t) \}^{-1}\sum_{l:T_l\geq t} \Bigg\{  I(T_l > t + \Delta t)\{1-\hat{\pi}_l(t, \Delta t) \}^2  +
\delta_l I(T_l < t + \Delta t)\{0-\hat{\pi}_l(t, \Delta t)\}^2  + \\
(1-\delta_l) I(T_l<t + \Delta t) \Big[  \hat{\pi}_l(T_l, \Delta t) \{ 1-\hat{\pi}_l(t, \Delta t) \}^2 +\{1-\hat{\pi}_l(T_l, \Delta t)\} \{0-\hat{\pi}_l(t, \Delta t)  \}^2 \Big]\Bigg\}

 \end{array}
\]
where $\mathcal{R}(t)$ denotes the number of subjects at risk at $t$. The term $I(T_l > t + \Delta t)\{1-\hat{\pi}_l(t, \Delta t) \}^2$ refers to patients who were alive after $t + \Delta t$ and $\delta_l I(T_l <t + \Delta t)\{0-\hat{\pi}_l(t, \Delta t)\}^2$ to patients who died before $t + \Delta t$. The remaining term refers to patients who were censored in the interval $[t, t + \Delta t]$.

\section{Analysis of the Cardio Data Set}\label{Analys}
In this section we present the analysis of the cardio data introduced in Section~\ref{Intro}. Our primary focus is to investigate the association between SAG with time-to-death/reoperation. 
In Figure~\ref{profAoG}, the evolution of the SAG for 12 randomly selected patients is presented, where it is shown that most of these patients have non-linear profiles. Therefore, we assumed a linear mixed-effects submodel including natural cubic splines for time. The DIC criterion indicated that the model assuming one internal knot at 5.02 years (corresponding to 50\% of the observed follow-up times) in both the fixed- and random-effects parts had a better fit. Furthermore, we corrected for gender. The mixed-effects submodel for the SAG is
\[
{y_{i}(t)}= \eta_{i}(t) + \epsilon_{i}(t)=
\beta_{0} +  \beta_{1} g_i+   \sum_{v=1}^2   \beta_{v+1}ns(t;v)  + b_{0i} + \sum_{v=1}^2  b_{vi}ns(t;v) +  \epsilon_{i}(t),
\]
where $y_{i}(t)$ are the measurements of SAG, $ns(.)$ denotes the natural cubic splines, $\epsilon_i(t) \sim N(0, \sigma)$ and $\boldsymbol{b}_i \sim N_b(\boldsymbol{0},\boldsymbol{\Sigma}_b^2)$.

To investigate the association between SAG and survival, we postulated the VCJM. Motivated by previous work \citep{andrinopoulou2015combined}, where different features of the SAG were found to have an influence on survival, we assumed the value and the slope of the longitudinal outcome to be associated with death/reoperation. Furthermore, we corrected for gender. Specifically, the survival submodel takes the form
\[
\begin{array}{ll}%
h_{i}(t,\boldsymbol{\theta}_s)&=\exp\Bigg\{\sum_{u=1}^{U}\gamma_{h_0,u}B_u(t)\Bigg\} \exp \Bigg\{\gamma g_i+ \lambda_1(t)\eta_i(t) + \lambda_2(t)\frac{\textrm{d}\eta_i(t)}{\textrm{d}t} \Bigg\}\\
&=\exp\Bigg\{\sum_{u=1}^{U}\gamma_{h_0,u}B_u(t)\Bigg\}\exp \Bigg\{\gamma g_i+ \sum_{\ell=1}^{L}\alpha_{1\ell} B_\ell(t)\eta_i(t) + \sum_{\ell=1}^{L}\alpha_{2\ell} B_\ell(t)\frac{\textrm{d}\eta_i(t)}{\textrm{d}t} \Bigg\},
 \end{array}
\]
where $\alpha_{1\ell}$ and $\alpha_{2\ell}$ are the coefficients that link the longitudinal and survival processes, $\gamma$ is the coefficient for gender and $\gamma_{h_0,\ell}$ are the baseline hazard coefficients. We assumed in both cases quadratic B-splines basis $B_\ell(t)=B_u(t)$ with 8 equally distance internal knots ranging from zero until 20.1 years.

As described in Section~\ref{Est}, for the P-splines approach we assumed normal priors for the time-varying coefficients $\boldsymbol{\alpha}= \{\boldsymbol{\alpha}_1,\boldsymbol{\alpha}_2\}$ and the baseline hazard coefficients $\boldsymbol{\gamma_{h_0}}$ and gamma hyperpriors for $\tau_{\alpha}$ and $\tau_{\gamma_{h_0}}$ where we took $c_1 = f_1 = 1$ and $c_2 = f_2 = 0.005$. For the rest of the parameters standard noninformative priors were used. For the coefficients of the longitudinal outcome $\boldsymbol{\beta}$ and the survival coefficient $\gamma$ normal priors were taken with mean zero and large variance. For the variance-covariance matrix of the random effects $\boldsymbol{\Sigma}_b$ we assumed inverse Wishart prior with an identity scale matrix and degrees of freedom equal to the total number of the random effects. For the precision parameter of the longitudinal outcome a gamma prior was taken with parameters that were based on the separate analysis of the outcome. We ran the MCMC using three chains with 150,000 iterations, 50,000 burn-in and 2 thinning.

\begin{figure}[h!]
\centerline{%
\includegraphics[height=3.5in]{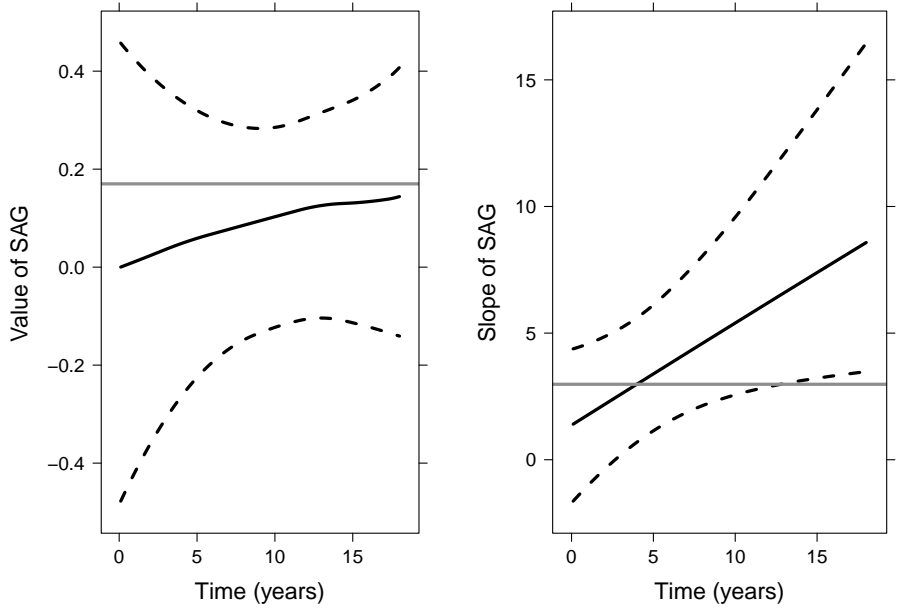}}
\caption{With black lines the mean estimates (solid line) and the credible intervals (dashed lines) of $\lambda_1(t)$ and $\lambda_2(t)$ functions corresponding to the value and slope association parameters are presented. The grey solid lines represent the mean estimates of the CCJM.}
\label{alphaDalpha}
\end{figure}

In Figure~\ref{alphaDalpha} we present the mean estimates and the credible intervals of the estimated $\lambda(t)$ for the value and the slope association parameters, respectively. The grey solid lines represent the coefficient when CCJM was assumed. The CCJM takes the form,
\[
h_{i}(t,\boldsymbol{\theta}_s)=\exp\Bigg\{\sum_{u=1}^{U}\gamma_{h_0,u}B_u(t)\Bigg\}
\exp \Bigg\{\gamma g_i+ \alpha_{1} \eta_i(t) + \alpha_{2} \frac{\textrm{d}\eta_i(t)}{\textrm{d}t} \Bigg\},
\]
where we use P-splines for the baseline hazard for a fair comparison. We observe that the effect of the SAG on survival seems to slightly increase over time, however this effect is not strong. The effect of the slope of SAG on survival appears to increase linearly with time. Specifically, at the beginning of the study ($t = 5$), for patients having the same gender and level of SAG, the log hazard ratio for one mmHg increase in the current slope of the SAG is 3.4. However, at the end of the study ($t = 15$) this effect increases to 7.4. Further results of the models including the posterior estimates, DIC and Gelman-Rubin's diagnostic are presented in Tables~\ref{ResultsTD} and~\ref{ResultsCon} and a discussion (Appendix). In Figure~\ref{pred10}, we present prediction plots for a 46 year old male. Every time the patient visits the hospital his survival/intervention-free probabilities are updated. Specifically, five years after his first visit his survival/intervention-free probability is 0.7, while five years after his last visit this probability is 0.3. The evolution of the SAG for the specific patient is presented in Figure~\ref{prof10}.

\begin{figure}[h!]
\centerline{%
\includegraphics[width=4.5in]{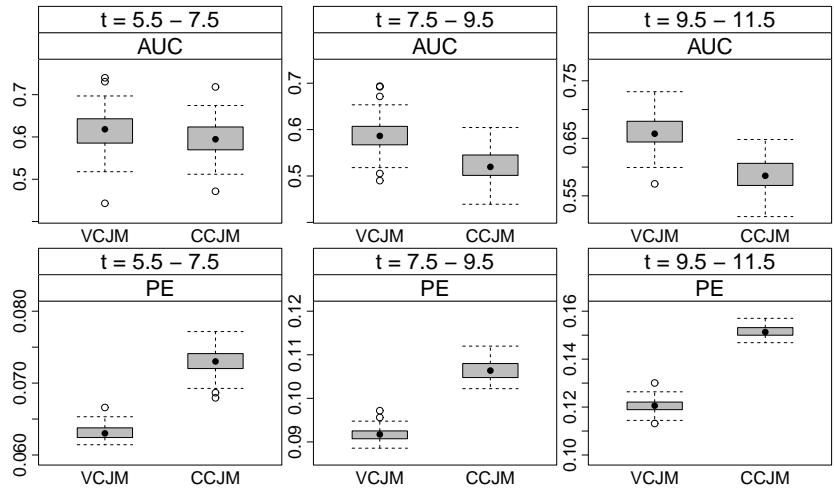}}
\caption{Boxplots of the AUC and PE measurements when assuming the VCJM and CCJM at different time points with $\Delta t = 2$ - Internal validation.}
\label{intVal1}
\end{figure}

To investigate whether the proposed VCJM improves dynamic predictions, we compared the VCJM with the CCJM based on the $\mbox{AUC}(t, \Delta t)$ and $\mbox{PE}(t, \Delta t)$ measures introduced in Section~\ref{predPerf}. Corrected estimates of these measures were obtained using an internal validation procedure. We performed a 5-fold cross-validation by splitting our data set in five subsets, fitting each time the model in four of the subsets and calculating the accuracy measures in the subset that was excluded. This cross-validation procedure was replicated 100 times. The calculation of $\mbox{AUC}(t, \Delta t)$ and $\mbox{PE}(t, \Delta t)$ was performed for the follow-up times $\{t = 5.5, 7.5, 9.5\}$ whereas $\Delta t = 2$. The results are presented in Figure~\ref{intVal1} where boxplots of 100 cross validations are shown. We obtained better discriminative capability (AUC) and predictive accuracy (PE) from the VCJM model compared to the CCJM in all cases. Specifically, we observe most of the time higher AUC values and lower PE values for the VCJM. These differences seem to be more profound for later follow-up times.

\section{Simulation study}\label{simul}

\subsection{Design} \label{SimDesign}
We performed a series of simulations to evaluate the performance of the VCJM. We simulated 400 patients with maximum number of repeated measurements equal to 10. We assumed one longitudinal and one survival outcome as in the analysis of the cardio data set. For the continuous longitudinal outcome, we investigated the following linear mixed-effects model
\[
y_{i}(t) = \eta_i(t) + \epsilon_{i}(t) = \beta_{0} +  \beta_{1} g_i + \beta_{2} t + b_{0i} + b_{1i}t+ \epsilon_{i}(t),
\]
where $\epsilon_i \sim N(0,\sigma_y^2)$ and $\boldsymbol{b}=(b_{0i}, b_{1i}) \sim N_2(\boldsymbol{0},\boldsymbol{\Sigma}_b^2)$. For simplicity, we adopted a linear effect of time for both the fixed and the random part, and corrected for gender. Time $t$ was simulated from a uniform distribution between zero and $19.5$.
For the survival part, we investigated the following scenarios.

For \textbf{Scenario I}, we postulated the following model:
\[
\begin{array}{ll}%
h_{i}(t)&=h_{0}(t)\exp \Bigg\{\gamma_{} g_i+ \lambda(t)\eta_i(t) \Bigg\}=h_{0}(t)\exp \Bigg\{\gamma_{} g_i+ \sum_{\ell=1}^{L}\alpha_{\ell} B_\ell(t)\eta_i(t) \Bigg\},
 \end{array}
\]
where $B_\ell(t)$ denotes the $\ell$-th basis function of a B-spline where the knots were placed at fixed time points. We assumed time-varying effect of the underlying value for the association parameter and corrected for gender. The baseline risk was simulated from a Weibull distribution $h_0(t)=\xi t^{\xi -1}$. For the simulation of the censoring times, an exponential censoring distribution was chosen with mean $\mu_{c}$, so that the censoring rate was between 40\% and 60\%. \textbf{Scenario Ia} assumed a linear evolution for $\lambda(t)$, while \textbf{Scenario Ib} a non-linear evolution.

For \textbf{Scenario II}, the survival submodel takes the form,
\[
h_{i}(t)=h_{0}(t)\exp \Bigg\{\gamma_{} g_i+ \alpha\eta_i(t) \Bigg\},
\]
where a constant effect for the $\alpha$ coefficient was assumed.

Finally, for \textbf{Scenario IIIa} and \textbf{Scenario IIIb}, the survival submodel takes the form,
\[
\begin{array}{ll}%
& h_{i}(t)=h_{0}(t)\exp \Bigg\{\gamma_{} g_i+ \alpha_1 + \alpha_2\eta_i(t) + \alpha_3\eta_i^2(t) \Bigg\}, \\
\textnormal{and} \ \ & h_{i}(t)=h_{0}(t)\exp \Bigg\{\gamma_{} g_i+ \alpha_1 + \alpha_2\eta_i(t) + \alpha_3\eta_i^2(t) +  \alpha_4\eta_i^3(t) +  \alpha_5\eta_i^4(t) \Bigg\},
 \end{array}
\]
respectively. In this case, we assumed polynomial effects for the association parameters.

We simulated 200 data sets per scenario. More details are presented in Table~\ref{Sim}.

\subsection{Analysis and Results}
To evaluate the performance of the proposed model, we fit the VCJM and the CCJM with the same specification as the one presented in Section~\ref{SimDesign}. To mimic the analysis of the cardio data set, the joint models were fitted using P-splines for the baseline hazard, rather than the Weibull one. Figure~\ref{sim8} and~\ref{sim20}, illustrate the true and estimated $\lambda(t)$ function when simulating from the linear Scenario Ia (left panel) and the non-linear Scenario Ib (right panel) and when fitting the VCJM. In particular, in Figure~\ref{sim8} we assumed 8 internal knots for the simulation and the fitting part while in Figure~\ref{sim20} we assumed 20. Figure~\ref{simCon} illustrates the true and estimated $\lambda(t)$ function for Scenario II when fitting the VCJM assuming 8 internal knots. Figures~\ref{simPoly1} and~\ref{simPoly2} present the results when simulating for Scenario IIIa and Scenario IIIb respectively and when fitting the VCJM with 8 internal knots. Overall our model successfully recovers the true $\lambda(t)$. The reason for observing greater variability at $t < 5$ and $t > 15$ is that relatively few events are observed in these regions.

\begin{figure}[h!]
\centerline{%
\includegraphics[width=4.5in]{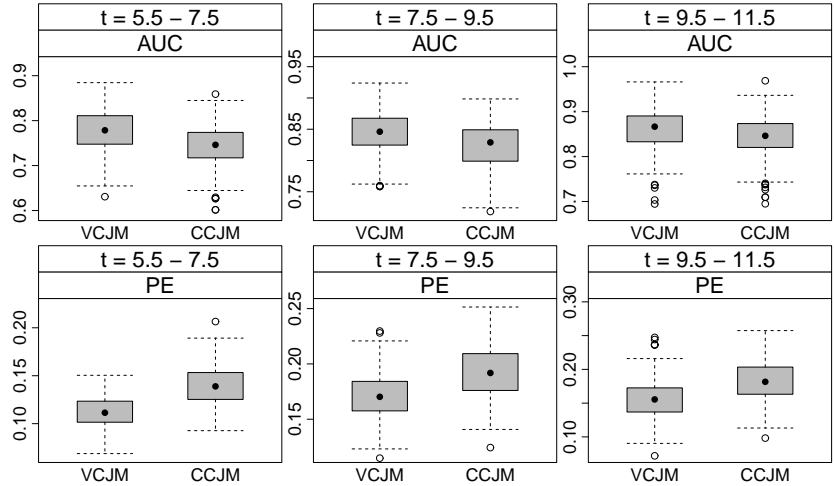}}
\caption{Boxplots of the AUC and PE measurements when assuming the VCJM and the CCJM at different time points with $\Delta t = 2$. For the simulation of the data scenario Ia was used - External validation.}
\label{simExValLin2Dt}
\end{figure}

\begin{figure}[h!]
\centerline{%
\includegraphics[width=4.5in]{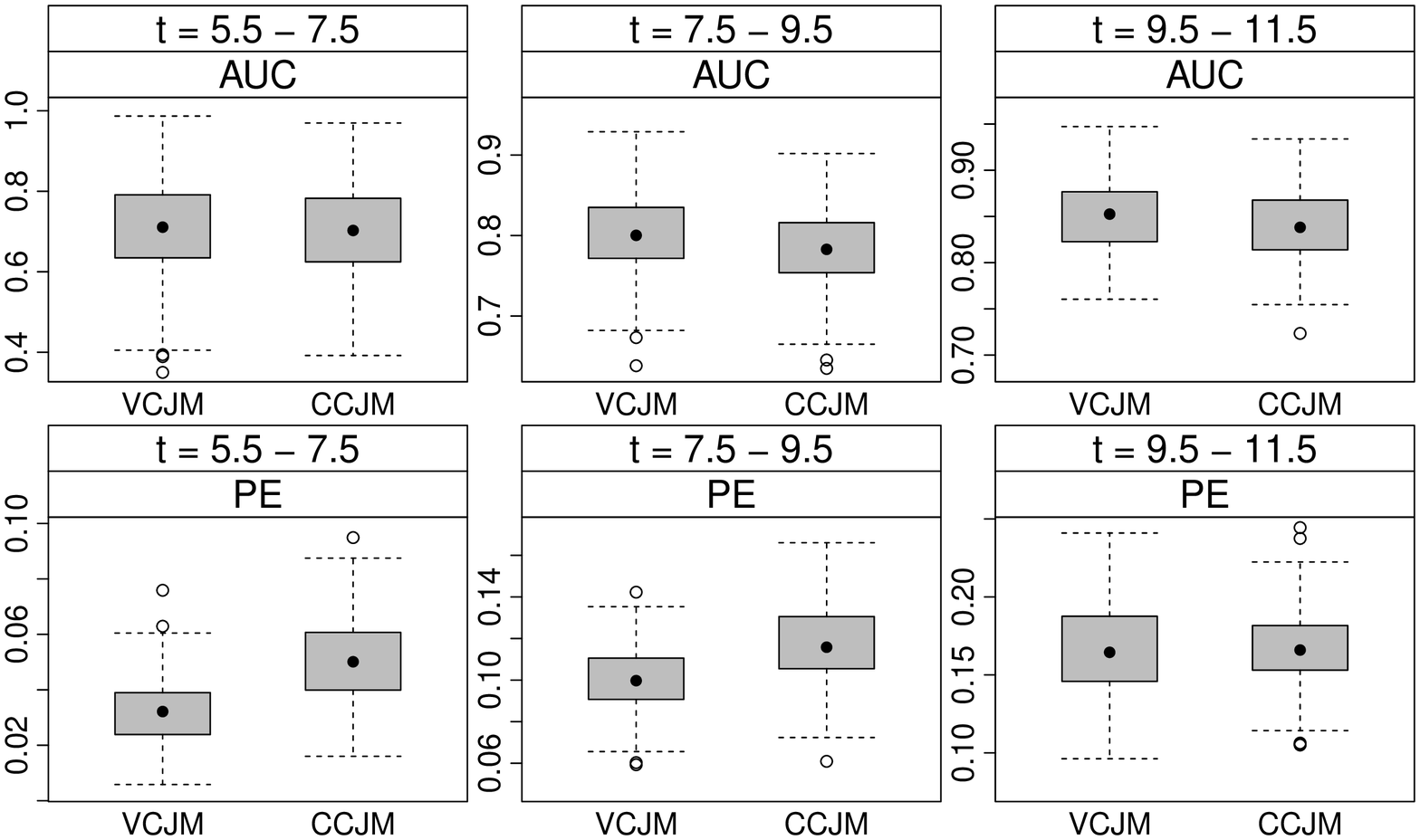}}
\caption{Boxplots of the AUC and PE measurements when assuming the VCJM and the CCJM at different time points with $\Delta t = 2$. For the simulation of the data scenario Ib was used - External validation.}
\label{simExValNonlin2Dt}
\end{figure}

\begin{figure}[h!]
\centerline{%
\includegraphics[width=4.5in]{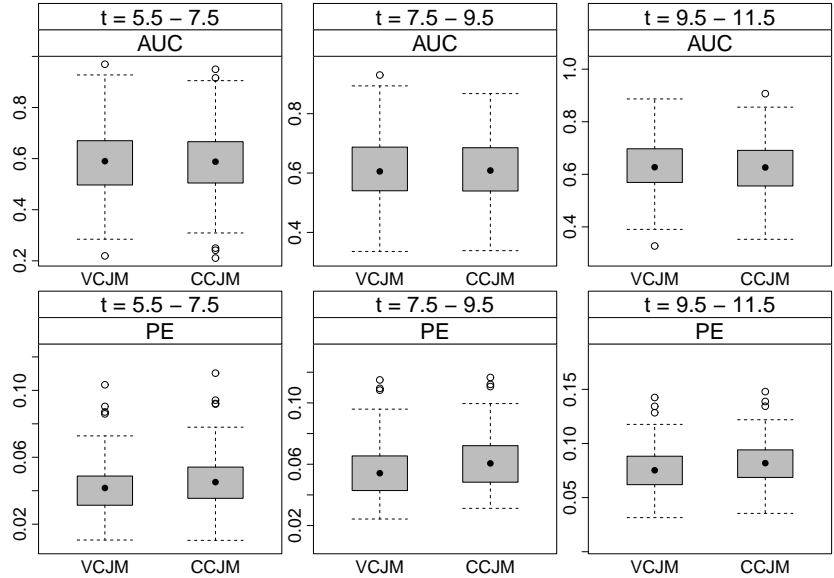}}
\caption{Boxplots of the AUC and PE measurements when assuming the VCJM and the CCJM at different time points with $\Delta t = 2$. For the simulation of the data scenario II was used - External validation.}
\label{simExValCon2Dt}
\end{figure}

To further evaluate whether the VCJM produces predictions of better quality than the CCJM, we performed an external validation procedure. For each simulated data set from scenarios Ia, Ib and II, we randomly excluded 200 patients. Using the remaining patients, we fitted the VCJM and the CCJM and computed the $\mbox{AUC}(t, \Delta t)$ and $\mbox{PE}(t, \Delta t)$ for the 200 patients that were initially excluded. These measures were calculated at follow-up times $\{t = 5.5, 7.5, 9.5\}$ using $\Delta t = 2$. Figures~\ref{simExValLin2Dt} and~\ref{simExValNonlin2Dt} present boxplots with the results under Scenarios Ia and Ib, respectively. It can be seen that, overall, the VCJM performs better than the CCJM. Most of the time we observe a higher AUC value and a lower PE value for the VCJM. Smaller differences are obtained for Scenario Ib for the AUC at $t = 5.5$ and the PE at $t = 9.5$. This is explained by the fact that not enough events were observed at the specific time periods. Figure~\ref{simExValCon2Dt}, which presents boxplots for Scenario II, suggests that the VCJM provides accurate predictions even if the data is generated with a constant effect for the coefficient that links the longitudinal and survival outcomes. In all cases, we obtain AUC and PE values that are similar in both the VCJM and the CCJM.

\section{Discussion}\label{Disc}
Motivated by the fact that the human tissue valves degenerate and hence the effect of SAG on survival could change over time, we developed a VCJM using P-splines. We used P-splines for the logarithm of the baseline hazard. We showed that even when the data is generated with a constant effect for the coefficient that links the longitudinal and survival outcome, the VCJM performs equal or better than the CCJM. In this paper we followed the Bayesian framework and therefore, extensions to more complex situations are comparably easy. A further advantage of the Bayesian approach, is that it allows us to automatically estimate the smoothing parameter in the P-splines approach by assigning a prior to it.

Further extensions to improve the VCJM can be considered for future work. In this paper we used only one longitudinal and one survival outcome. Extension to multiple longitudinal outcomes might be useful since other longitudinal outcomes that are related to the heart value may have an influence on survival. Finally, death and reoperation are competing risks and therefore a competing risk survival submodel would be more appropriate.

\bibliographystyle{abbrvnat} 
\bibliography{mybiblo}

\newpage

\section*{Appendix}
\appendix

\subsection*{Likelihood and DIC}

The likelihood contribution of the longitudinal submodel takes the form
\begin{eqnarray*}
	p(y_{ij}\mid \boldsymbol{\theta}_{y}, \boldsymbol{b}_{i})=
	(2\pi \sigma)^{-1/2} \exp{\biggl[  -\frac{(y_{ij} -\boldsymbol{x_{ij}}^\top\boldsymbol{\beta} - \boldsymbol{z_{ij}}^\top\boldsymbol{b_i})^2}{2\sigma^2}    \biggr] },
\end{eqnarray*}
where $\boldsymbol{\theta}_{y}$ are the parameters of the longitudinal submodel. The likelihood contribution of the survival submodel is given by
\begin{eqnarray*}
	\lefteqn{p\{T_i,\delta_i\mid \mathcal \eta_{i}(T_i),\boldsymbol{\theta}_s\}= } \\
	&&\exp\biggr[\sum_{q=1}^Q\gamma_{h_{0},q}B_q(T_i,\boldsymbol{\nu})+ \gamma^{\top}\boldsymbol{w}_{i}+ f\{ \sum_{l=1}^L\alpha_l B_l(T_i), \eta_i(T_i)\} \biggr] ^{I(\delta_i=1)}\times \\
	&& \exp{\biggl\{-\exp{(\boldsymbol{\gamma}^{\top} \boldsymbol{w}_{i})\int_0^{T_i}} \exp{\biggl[\sum_{q=1}^Q \gamma_{h_{0k},q}B_q(s,\boldsymbol{\nu})+ f\{\sum_{l=1}^L\alpha_l B_l(s), \eta_i(s)\}\biggr]}ds  \biggr\}},
\end{eqnarray*}
where $\boldsymbol{\theta}_{s}$ are the parameters of the survival submodel. The integral of the survival function does not have a closed-form solution, and thus a numerical method must be employed for this evaluation. To approximate this integral we used the Gaussian quadrature rule and we assume a 15-point Gauss-Kronrod rule.

The joint likelihood function is written as
\begin{eqnarray*}
	p(\boldsymbol{y}_{i},T_i,\delta_i \mid  \boldsymbol{\theta}, \boldsymbol{b}_i) =   \prod_{j=1}^{n_{i}}p(y_{ij}\mid \boldsymbol{b}_{i},\boldsymbol{\theta}_{y})
	p\{T_i,\delta_i\mid \mathcal \eta_{i}(T_i),\boldsymbol{\theta}_s\}p(\boldsymbol{b}_{i}\mid \boldsymbol{\theta}_{y}),
\end{eqnarray*}
where $\boldsymbol{\theta} = (\boldsymbol{\theta}_s^{\top}, \boldsymbol{\theta}_{y}^{\top})^{\top}$ and
\begin{eqnarray*}
	p(\boldsymbol{b}_{i}\mid \boldsymbol{\theta}_{y}) = [2\pi\det(\boldsymbol{D})]^{-1/2}\exp{\biggr(- \frac{\boldsymbol{b}_i^\top \boldsymbol{D}^{-1} \boldsymbol{b_i} }{2} \biggr) }.
\end{eqnarray*}

We use the DIC constructed from the conditional distribution where the random effects are treated as parameters since it facilitates computations. In particular, the DIC takes the form
\begin{eqnarray*}
	DIC = pD + \bar{D},
\end{eqnarray*}
where
\begin{eqnarray*}
	pD &=& \bar{D} - D(\bar{\boldsymbol{\theta}}, \bar{\boldsymbol{b}}),\\
	\bar{D}  &=& E\{D[\boldsymbol{\theta}^{(g)}, \boldsymbol{b}^{(g)}]\},\\
	D[\boldsymbol{\theta^{(g)}}, \boldsymbol{b}^{(g)}] &=& -2 \sum_{i=1}^n log\{p[\boldsymbol{y}_{i},T_i,\delta_i \mid \boldsymbol{\theta}^{(g)}, \boldsymbol{b}^{(g)}]\},
\end{eqnarray*}
where $\boldsymbol{b} = {\boldsymbol{b}_1, \dots , \boldsymbol{b}_n}$ are the random effects, $g=1, \dots, G$ is the iteration of the sampler, $\boldsymbol{\theta}^{(g)}$ and $\boldsymbol{b}^{(g)}$ denote the parameter samples at the \textit{g}th iteration and $\bar{\boldsymbol{\theta}}$ and $\bar{\boldsymbol{b}}$ represent the means of the posterior samples. The model with the smaller DIC value represents the model with the best fit. 

\subsection*{Discussion of the data results}
In Table~\ref{ResultsTD} and ~\ref{ResultsCon} we present the posterior means, the standard errors, the 95\% credible intervals and the Gelman-Rubin's Rhat diagnostic of the VCJM and CCJM respectively. 

With regards to the VCJM, it can be seen that gender does not seem to be a strong factor for the SAG. A nonlinear effect of time seems to capture the evolution of SAG. Furthermore, gender does not seem to have an influence on the survival outcome. The results of the association parameters are difficult to interpret therefore plots were created in the manuscript (Figure 2 in the main paper) where it can be seen that the slope of the SAG on survival appears to increase linearly with time. The Gelman-Rubin's diagnostic (potential scale reduction factor) suggests that all the parameters of the VCJM converged.

With regards to the CCJM, we found the same results for the longitudinal submodel as in the VCJM. Moreover, gender does not seem to have an influence on the survival outcome. The underlying value and slope of the SAG seem to have an effect on the survival outcome. In particular, the log hazard is increased by 0.17, with 95\% credible interval [0.03, 0.30], for each unit increase in the current value of the SAG. Furthermore, the log hazard is increased by 2.98, with 95\% credible interval [1.45, 4.75], for each unit increase in the current slope of the SAG. The Gelman-Rubin's diagnostic suggests that all the parameters of the CCJM converged.

The DIC values indicate that the VCJM provides a better fit of the data compared to the CCJM.

\counterwithin{figure}{section}
\section{Figures}

\begin{figure}[htp]
\centerline{%
\includegraphics[height=4in]{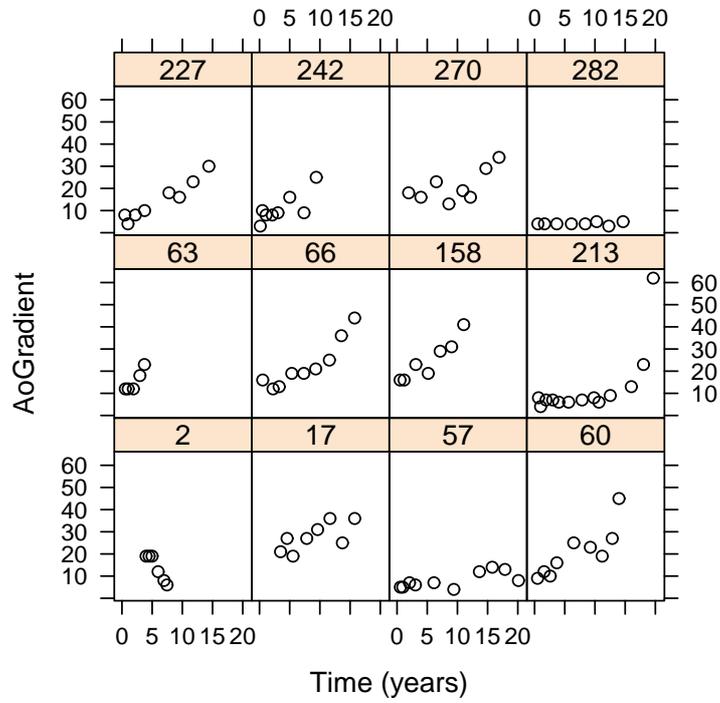}}
\caption{Profile of the SAG for 12 randomly selected patients.}
\label{profAoG}
\end{figure}

\newpage

\begin{figure}[h!]
\centerline{%
\includegraphics[width=6in]{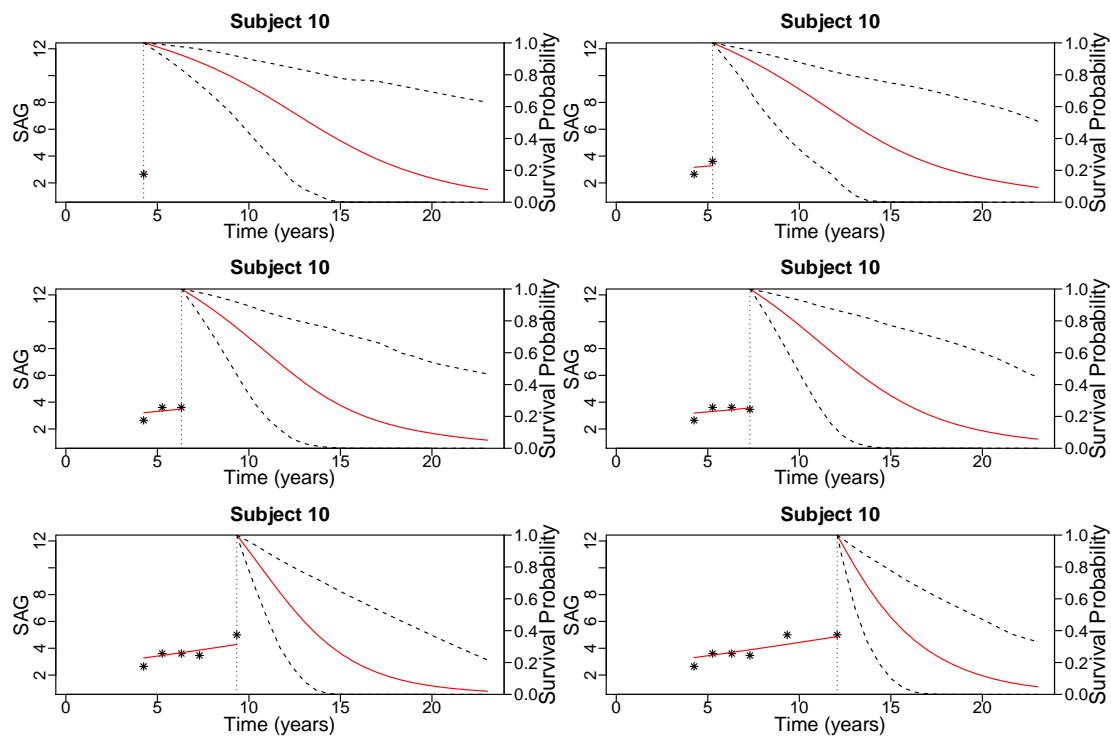}}
\caption{Dynamic survival/intervention-free predictions for patient 10. The vertical dotted lines represent the time point of the last SAG measurement.  On the left side, the fitted longitudinal trajectory is presented. On the right side, the solid line represents the mean estimator of the predictions while the dashed lines the corresponding 95\% confidence interval.}
\label{pred10}
\end{figure}

\newpage

\begin{figure}[h!]
\centerline{%
\includegraphics[width=5in]{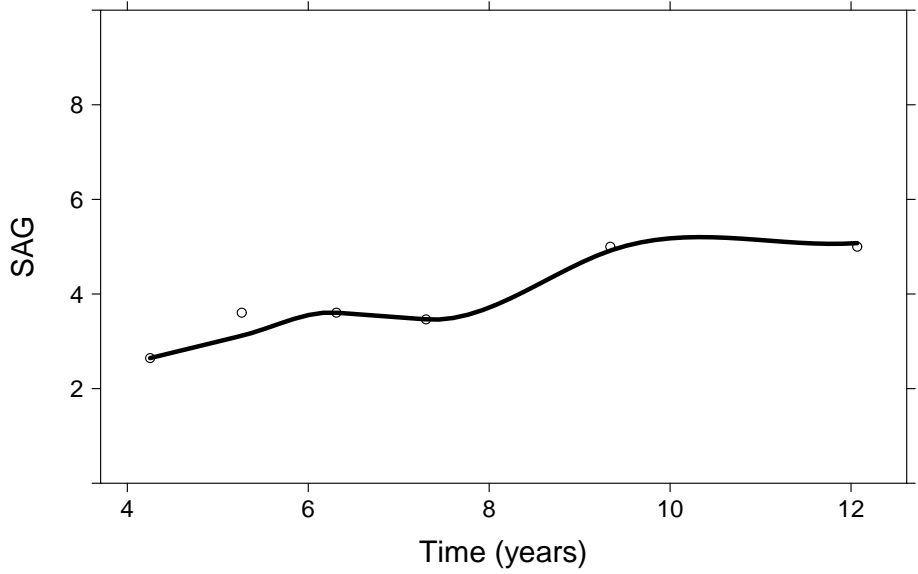}}
\caption{The SAG evolution for patient 10.}
\label{prof10}
\end{figure}

\begin{figure}[h!]
\centerline{%
\includegraphics[height=3.5in]{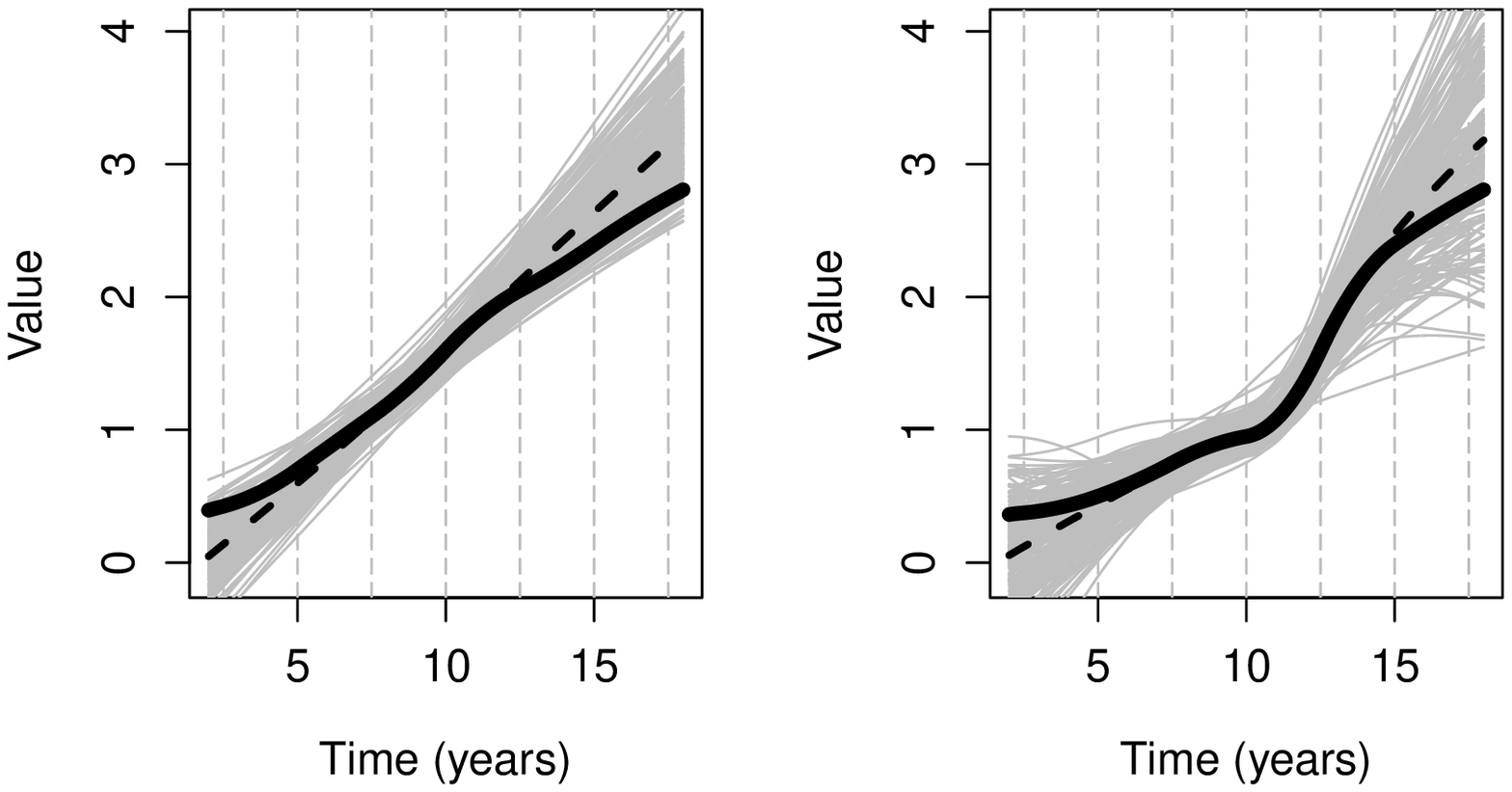}}
\caption{Simulation results assuming a time-varying effect (Scenario I) and 8 internal knots when simulating and assuming the VCJM with the same knots when fitting the data: the left panel corresponds to Scenario Ia (linear $\lambda(t)$) and the right panel to Scenario Ib (non-linear $\lambda(t)$). The solid black lines denote the true $\lambda(t)$ function, the dashed black lines the average estimates of this function from the 200 data sets and the grey lines the estimates from each of the 200 data sets.}
\label{sim8}
\end{figure}

\newpage

\begin{figure}[h!]
\centerline{%
\includegraphics[height=3.5in]{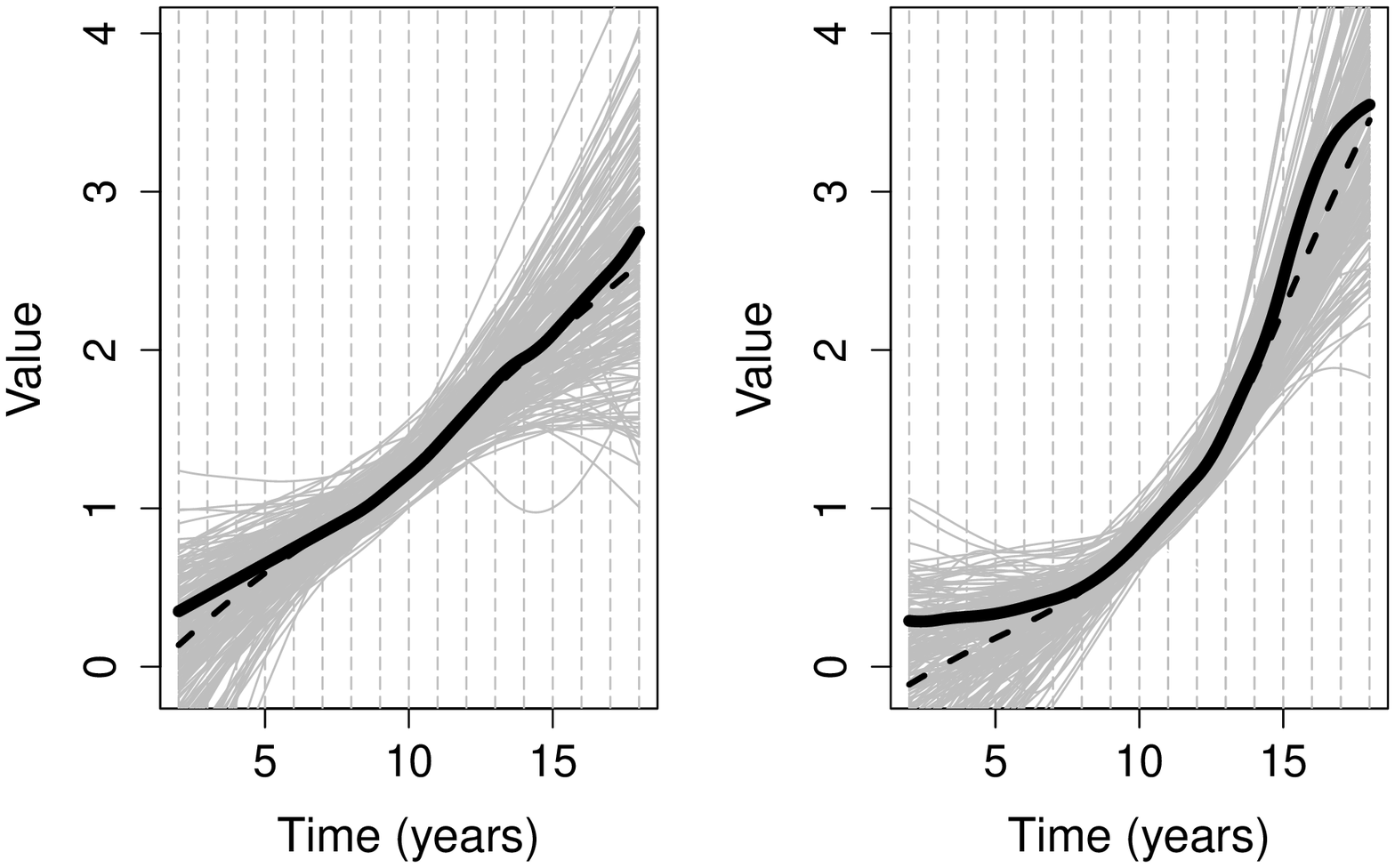}}
\caption{Simulation results assuming a time-varying effect (Scenario I) and 20 internal knots when simulating and assuming the VCJM with the same knots when fitting the data: the left panel corresponds to Scenario Ia (linear $\lambda(t)$) and the right panel to Scenario Ib (non-linear $\lambda(t)$). The solid black lines denote the true $\lambda(t)$ function, the dashed black lines the average estimates of this function from the 200 data sets and the grey lines the estimates from each of the 200 data sets.}
\label{sim20}
\end{figure}

\newpage

\begin{figure}[h!]
\centerline{%
\includegraphics[height=3.5in]{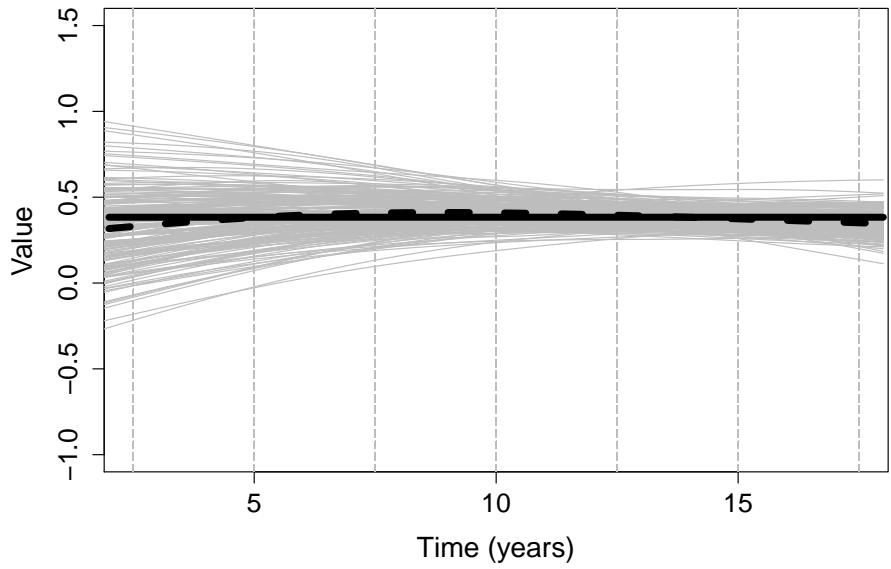}}
\caption{Simulation results assuming a constant effect (Scenario II) when simulating and assuming the VCJM with 8 internal knots when fitting the data: The solid black lines denote the true association parameter, the dashed black lines the average estimates of this function from the 200 data sets and the grey lines the estimates from each of the 200 data sets.}
\label{simCon}
\end{figure}

\newpage

\begin{figure}[h!]
\centerline{%
\includegraphics[height=3.5in]{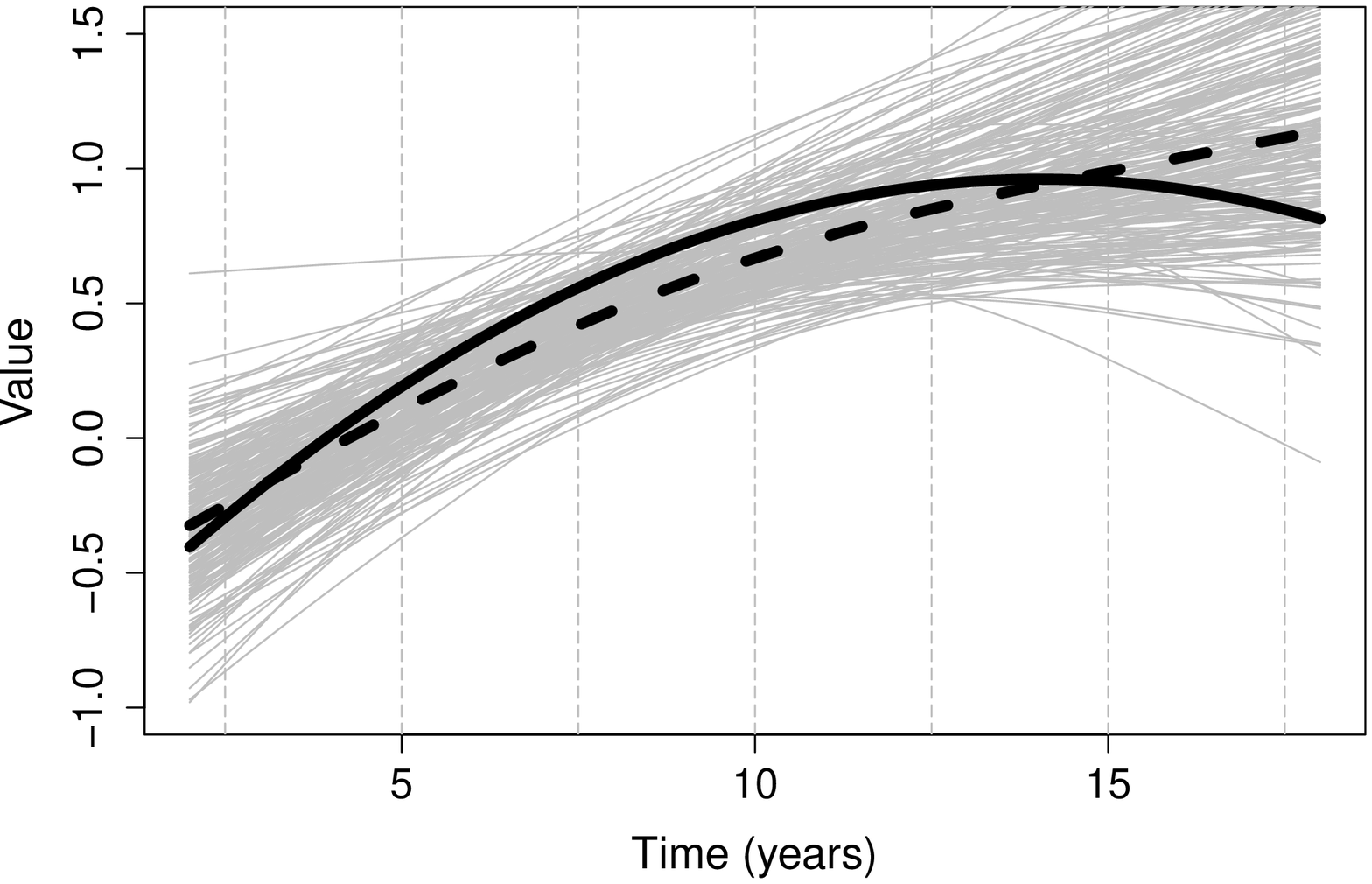}}
\caption{Simulation results assuming a second degree polynomial for the time function of the time-varying coefficient (Scenario IIIa) when simulating and assuming the VCJM with 8 internal knots when fitting the data: The solid black lines denote the true association parameter, the dashed black lines the average estimates of this function from the 200 data sets and the grey lines the estimates from each of the 200 data sets.}
\label{simPoly1}
\end{figure}

\newpage

\begin{figure}[h!]
\centerline{%
\includegraphics[height=3.5in]{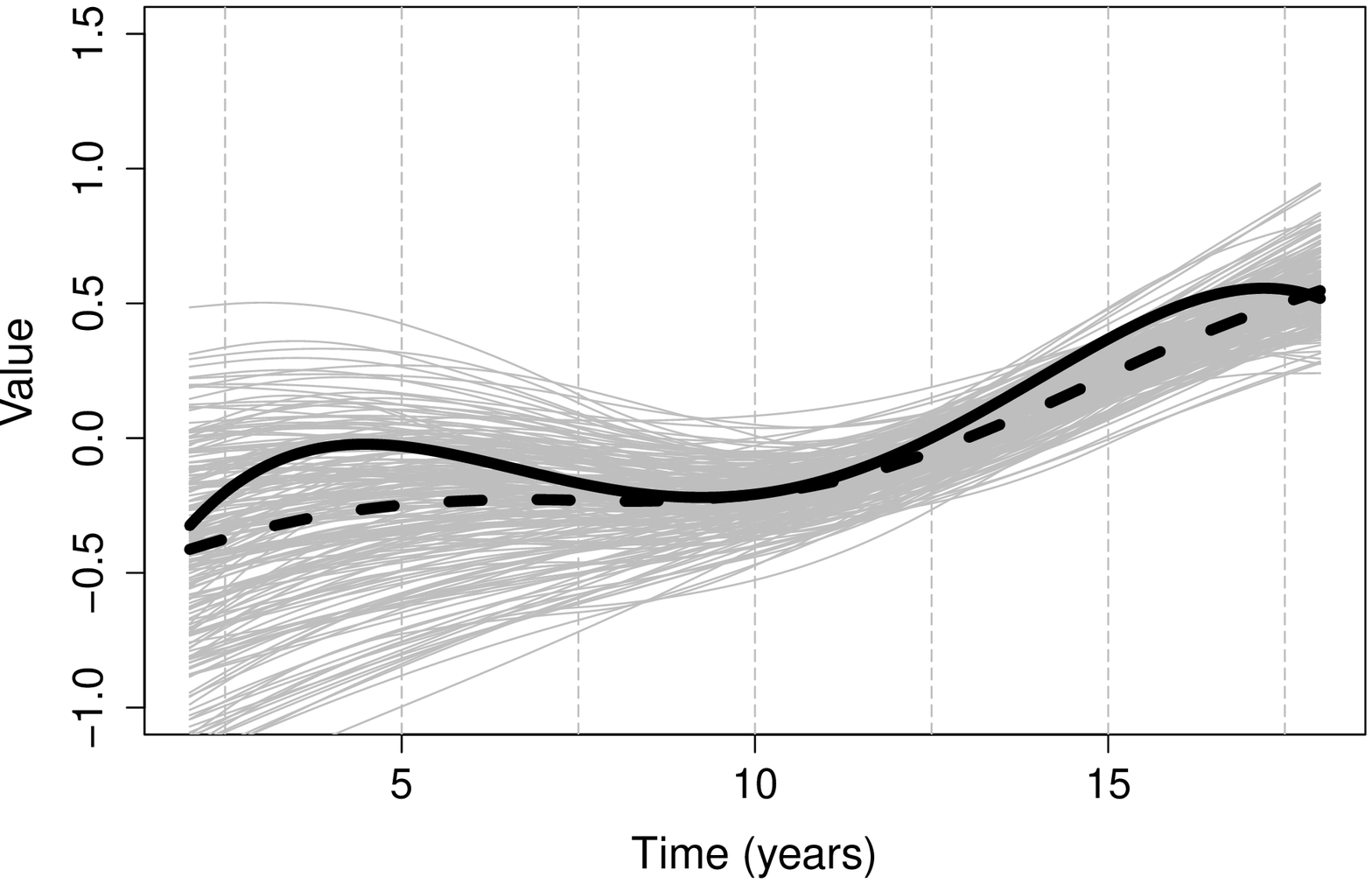}}
\caption{Simulation results assuming a fourth degree polynomial for the time function of the time-varying coefficient (Scenario IIIb) when simulating and assuming the VCJM with 8 internal knots when fitting the data: The solid black lines denote the true association parameter, the dashed black lines the average estimates of this function from the 200 data sets and the grey lines the estimates from each of the 200 data sets.}
\label{simPoly2}
\end{figure}

\newpage

\counterwithin{table}{section}

\section{Tables}

\begin{table}[ht]
\caption{VCJM results: posterior means, standard errors, 95\% credible intervals and Rubin and Gelman-Rubin's diagnostic. Rhat = potential scale reduction factor.}
\centering
\begin{tabular}{rrrrrr}
  \hline
& Mean & SE & 2.5\% & 97.5\%  & Rhat\\
\hline 
\it{SAG} &&&&\\
  \hline
  (intercept) & 3.02 & 0.00 & 2.87 & 3.17 & 1.00 \\ 
  SexFemale & 0.18 & 0.00 & -0.07 & 0.44 & 1.00 \\ 
  ns(echotime, 2)1 & 3.83 & 0.01 & 3.29 & 4.38 & 1.00 \\ 
  ns(echotime, 2)2 & 3.42 & 0.01 & 2.77 & 4.12 & 1.00 \\ 
\hline 
\it{Death/reoperation} &&&&\\
  \hline
  SexFemale & 0.01 & 0.00 & -0.38 & 0.38 & 1.00 \\ 
  alpha value[1] & -0.02 & 0.01 & -0.58 & 0.52 & 1.00 \\ 
  alpha value[2] & 0.02 & 0.01 & -0.39 & 0.40 & 1.00 \\ 
  alpha value[3] & 0.05 & 0.01 & -0.24 & 0.33 & 1.00 \\ 
  alpha value[4] & 0.08 & 0.00 & -0.16 & 0.29 & 1.00 \\ 
  alpha value[5] & 0.10 & 0.00 & -0.12 & 0.28 & 1.00 \\ 
  alpha value[6] & 0.13 & 0.00 & -0.10 & 0.32 & 1.00 \\ 
  alpha value[7] & 0.13 & 0.01 & -0.12 & 0.35 & 1.00 \\ 
  alpha value[8] & 0.15 & 0.01 & -0.15 & 0.42 & 1.00 \\ 
  alpha value[9] & 0.21 & 0.01 & -0.17 & 0.59 & 1.00 \\ 
  alpha value[10] & 0.29 & 0.01 & -0.21 & 0.85 & 1.00 \\ 
  alpha slope[1] & 0.78 & 0.05 & -2.74 & 4.09 & 1.00 \\ 
  alpha slope[2] & 1.95 & 0.04 & -0.67 & 4.63 & 1.00 \\ 
  alpha slope[3] & 3.11 & 0.04 & 0.91 & 5.67 & 1.00 \\ 
  alpha slope[4] & 4.27 & 0.06 & 1.97 & 7.42 & 1.00 \\ 
  alpha slope[5] & 5.42 & 0.08 & 2.60 & 9.59 & 1.00 \\ 
  alpha slope[6] & 6.57 & 0.10 & 3.01 & 12.03 & 1.00 \\ 
  alpha slope[7] & 7.71 & 0.12 & 3.30 & 14.52 & 1.00 \\ 
  alpha slope[8] & 8.85 & 0.15 & 3.54 & 17.06 & 1.00 \\ 
  alpha slope[9] & 9.99 & 0.17 & 3.68 & 19.59 & 1.00 \\ 
  alpha slope[10] & 11.13 & 0.19 & 3.82 & 22.20 & 1.00 \\ 
   \hline
   DIC & & & & & 8829.617\\
\end{tabular}
\label{ResultsTD}
\end{table}

\newpage

\begin{table}[ht]
\caption{CCJM results: posterior means, standard errors, 95\% credible intervals and Rubin and Gelman-Rubin's diagnostic. Rhat = potential scale reduction factor}
\centering
\begin{tabular}{rrrrrr}
  \hline
& Mean & SE & 2.5\% & 97.5\% & Rhat \\
\hline 
\it{SAG} &&&&\\
  \hline 
  (Intercept) & 3.01 & 0.00 & 2.86 & 3.16 & 1.00 \\ 
  SexFemale & 0.19 & 0.00 & -0.07 & 0.44 & 1.00 \\ 
  ns(echotime, 2)1 & 3.81 & 0.01 & 3.31 & 4.33 & 1.00 \\ 
  ns(echotime, 2)2 & 3.29 & 0.01 & 2.65 & 4.01 & 1.00 \\
\hline 
\it{Death/reoperation} &&&&\\
  \hline
  SexFemale & 0.01 & 0.00 & -0.36 & 0.36 & 1.00 \\ 
  alpha value & 0.17 & 0.00 & 0.03 & 0.30 & 1.00 \\ 
  alpha slope & 2.98 & 0.02 & 1.45 & 4.75 & 1.00 \\ 
   \hline
   DIC & & & & & 8917.099\\ 
\end{tabular}
\label{ResultsCon}
\end{table}

\newpage

{\small{
\begin{table}[!h]
\caption{Simulation scenarios. \label{Table1}}
\begin{center}
\begin{tabular}{lccccccl}
\hline
Scenario & $\boldsymbol{\beta}$ & $\boldsymbol{\sigma}_y$ & $diag\{\boldsymbol{\Sigma}_b\}$ & $\xi$ &$\mu_c$&$\boldsymbol{\gamma}$ & $\boldsymbol{\alpha}$  \\
\hline
Ia & && & & \\
\hline
& (Intercept) = 3.03 & 0.69 & 0.93 & 1.2&30 & (Intercept) = -7.85 & 0.19 \\
& Females = 0.14 & & 0.16 & && Females =-0.02 &  0.35\\
& Time = 0.16 &&& & & &0.5\\
& && & && &0.9\\
& && & && &1.3\\
& && &&&  &1.9\\
& && & && &2.2\\
& && & && &2.59\\
& && & && & 2.9\\
& && & && & 3.19\\
\hline
Ib  &&&& & & \\
\hline
& (Intercept) = 3.03 & 0.69 & 0.93 & 1.2& 30&(Intercept) = -7.75 & 0.19 \\
& Females = 0.14 & & 0.16 & && Females =-0.02 &  0.35\\
& Time = 0.16 &&&&&  &0.4\\
& && &&&  &0.6\\
& && &&&  &0.9\\
& && &&&  &1\\
& && &&&  &2.2\\
& && &&&  &2.59\\
& && & && &2.9\\
& && && & & 3.19\\
\hline
II  & && && & \\
\hline
& (Intercept) = 3.03 & 0.69 & 0.93& 1.9 &24& (Intercept) = -7.85 & 0.38 \\
& Females = 0.14 & & 0.16 & && Females =-0.02 &  \\
& Time =  0.16 &&& &&  &\\
\hline
IIIa &&&& & & \\
\hline
& (Intercept) = 3.02 & 0.69 & 0.93 & 1.7& 12&(Intercept) = -5.75 & -0.89 \\
& Females = 0.16 & & 0.16 & && Females = 0.01 &  0.26\\
& Time = 0.16 &&&&&  &-0.01\\
\hline
IIIb  &&&& & & \\
\hline
& (Intercept) = 3.02 & 0.69 & 0.93 & 1.7& 12&(Intercept) = -5.75 & -1.33 \\
& Females = 0.16 & & 0.16 & && Females = 0.01 &   0.75\\
& Time = 0.16 &&&&&  &-0.15\\
& && &&&  &0.01\\
& && &&&  &-0.0002\\
\label{Sim}
\end{tabular}
\end{center}
\end{table}
}}

\end{document}